\documentclass{bmcart}

\usepackage{amsthm,amsmath}
\usepackage{amsfonts}
\usepackage{fixltx2e}
\RequirePackage{natbib}
\RequirePackage{hyperref}
\usepackage[utf8]{inputenc} 
\usepackage{graphicx}

\startlocaldefs

\def\bra#1{{\langle #1 |}}
\def\ket#1{{| #1 \rangle}}
\endlocaldefs

\begin{document}

\begin{frontmatter}

\begin{fmbox}
\dochead{Research}


\title{Minimal ancilla mediated quantum computation}


\author[
   addressref={aff1},                   
   corref={aff1},                       
   email={py08tjp@leeds.ac.uk}   
]{\inits{TJ}\fnm{Timothy J} \snm{Proctor}}
\author[
    addressref={aff1},                   
    corref={aff1},                       
   email={v.kendon@leeds.ac.uk}
]{\inits{V}\fnm{Viv} \snm{Kendon}}


\address[id=aff1]{
  \orgname{School of Physics and Astronomy, E C Stoner Building, University of Leeds} 
  \street{},                     %
  \postcode{LS2 9JT}                                
  \city{Leeds},                              
  \cny{UK}                                    
}



\end{fmbox}


\begin{abstractbox}

\begin{abstract} 
Schemes of universal quantum computation in which the interactions between the computational elements, in a computational register, are mediated by some ancillary system are of interest due to their relevance to the physical implementation of a quantum computer. Furthermore, reducing the level of control required over both the ancillary and register systems has the potential to simplify any experimental implementation. In this paper we consider how to minimise the control needed to implement universal quantum computation in an ancilla-mediated fashion. Considering computational schemes which require no measurements and hence evolve by unitary dynamics for the global system, we show that when employing an ancilla qubit there are certain fixed-time ancilla-register interactions which, along with ancilla initialisation in the computational basis, are universal for quantum computation with no additional control of either the ancilla or the register. We develop two distinct models based on locally inequivalent interactions and we then discuss the relationship between these unitary models and the measurement-based ancilla-mediated models known as ancilla-driven quantum computation.

\end{abstract}


\begin{keyword}
\kwd{ancilla}
\kwd{universal gates}
\kwd{minimal control}
\kwd{quantum computation}
\kwd{quantum bus}
\kwd{ancilla-driven}
\kwd{ancilla-controlled}
\end{keyword}


\end{abstractbox}
%

\end{frontmatter}



\section{Introduction}
The original theoretical setting for quantum computation is the gate model \cite{feynman1985quantum} in which a global unitary that acts on a register of qubits, which computes the solution to some problem, is decomposed into a sequence of fundamental gates that are applied to the register. As in classical computation, it is desirable for these fundamental gates to be members of some \emph{finite} and \emph{universal} gate set, from which any global unitary can be composed up to arbitrary accuracy. There has been extensive research on such universal sets, and a significant example is the set composed of any entangling gate in conjunction with any universal set of single-qubit unitaries \cite{brylinski2002universal,bremner2002practical}. Furthermore, almost any two-qubit entangling gate is universal on its own provided that it can be applied to arbitrary pairs of qubits \cite{lloyd1995almost,barenco1995universal}. These results are of significant theoretical importance for the understanding of quantum computation.
\newline
\indent
However, the physical implementation of these models requires direct interactions between arbitrary pairs of  register qubits and, often, direct application of single-qubit rotations and measurements. This is a huge practical challenge and most experimentally implemented or proposed schemes mediate the required multi-qubit interactions using some ancillary system. An example of such an ancilla-mediated scheme is the original Cirac-Zoller ion-trap gate \cite{cirac1995quantum}, where the ancilla in this case is the collective quantized motion of the ions. Further examples include superconducting qubits coupled to nitrogen-vacancy (NV) centres \cite{marcos2010coupling,lu2013quantum,qiu2014coupling,zhu2011coherent} or transmission line resonators \cite{wang2009coupling,wallraff2004strong}, spin qubits coupled via ancillary photonic qubits \cite{carter2013quantum,luxmoore2013interfacing} and the coupling of a Cooper-pair box to a micro-mechanical resonator \cite{armour2002entanglement}. It is therefore of both practical and theoretical interest to study the effect of incorporating the ancillary system into the computational model. Indeed, those gates that have been shown to be universal in a direct implementation of the circuit model cannot in general be utilised to implement quantum computations entirely mediated via an ancilla. We will refer to schemes in which all the multi-qubit interactions are mediated via some ancillary system as \emph{ancilla-mediated quantum computation} (AMQC). 
\newline
\indent
An extensively researched model of AMQC is \emph{quantum bus} (qubus) computation \cite{milburn1999simulating,wang2002simulation,spiller2006quantum, brown2011ancilla, louis2007efficiencies,munro2005efficient}. This model employs a field-mode ancilla to mediate two-qubit gates on pairs of register qubits with the interaction between the ancilla and a register qubit being a controlled displacement of the field-mode.  Recently, we have developed an analogous model which employs a $d$-dimensional \emph{qudit} ancilla \cite{proctor2014quantum} with a displacement operator defined in the discrete phase space of the qudit \cite{wootters1987wigner,vourdas2004quantum}. These models have been shown to require a lower number of operations to implement certain gate sequences than a direct implementation of the circuit model \cite{brown2011ancilla,proctor2014quantum,noteE}. However, neither of these models can implement a universal gate set on the register using \emph{only} this ancilla-register interaction and so, although no interactions between register qubits are required, some further direct access is needed to the register qubits to implement some basis-changing single-qubit unitary \cite{proctor2014quantum,brown2011using}.
\newline
\indent
In order to implement useful quantum computations, it will be necessary to have register qubits with as long a coherence time as possible. However, if it is necessary to access each register qubit to implement multiple forms of control this will potentially introduce many sources of decoherence. Limiting the forms of access required to the register qubits may help to isolate the register more effectively, and so, motivated by this, the measurement-based \emph{ancilla-driven quantum computation} (ADQC) \cite{kashefi2009twisted,anders2010ancilla,shah2013ancilla} and more recently the globally unitary \emph{ancilla-controlled quantum computation} (ACQC) \cite{proctor2013universal} schemes were developed. In both of these models the access to the register qubits is limited to one \emph{fixed-time} interaction between a single register qubit (at a time) and an ancilla qubit, where the qubits are not necessarily of the same physical type. The additional direct access to the register that was required in the qubus architecture, and analogous qudit model, is replaced by local unitary operations on the ancilla and, in the case of ADQC, ancilla measurements.   
\newline
\indent
Although both the ancilla-driven and ancilla-controlled models require the minimum possible access to the computational register, they replace the local control of the register with local control of the ancillary system and so still require more than one fixed quantum gate to implement universal computation. Halil-Shah and Oi \cite{halil2014minimum} have recently shown that the measurement-based ancilla-driven model can be adapted so that no local control, beyond ancilla preparation in a fixed state, is required of either the ancilla or the register. In this model, the computation can be achieved using a fixed interaction and ancilla measurement in the computational basis alone. However, this requires a stochastic repeat-until-success style gate scheme \cite{paetznick2013repeat}, whereby one has to wait until a random walk through the set of unitaries is within the required precision of the desired unitary. In this paper we will show that it is possible to develop deterministic models that require only a single fixed ancilla-register interaction and ancilla preparation in the computational basis with no ancilla measurements necessary. Such schemes require a minimal level of control of both the ancillary and register systems whilst allowing for universal quantum computation. Hence, we will refer to such models as \emph{minimal control} models of ancilla-mediated quantum computation, and we will often drop the reference to ancilla-mediation for brevity. In Section~\ref{min1} we introduce two such schemes based on locally inequivalent interactions. The first of these models requires multiple ancillas to implement entangling two-qubit gates on the register and so, although it requires minimal control, it has an overhead in terms of ancilla use. We then develop an alternative minimal control model which does not have this ancilla overhead and requires only three ancilla-register interactions per two-qubit entangling gate, the minimum possible in any unitary scheme \cite{lamata2008sequential}. We briefly discuss the physical implementation of these models before concluding in Section~\ref{con}. We begin in Section~\ref{def} with some essential definitions. 

\section{Definitions \label{def}}
We denote the Pauli operators acting on the $j^{th}$ qubit by $X_j$, $Y_j$ and $Z_j$ and take $\ket{0}$ and $\ket{1}$ to be the positive and negative eigenstates of the Pauli $Z$ operator respectively. Using standard definitions, we take the Hadamard gate $H$ to be
\begin{equation} H := \frac{1}{\sqrt{2}}\left(\ket{0}\bra{0} + \ket{0}\bra{1} + \ket{1}\bra{0} - \ket{1}\bra{1} \right),\label{Had} \end{equation}
and the single-qubit phase gate to be
\begin{equation}R(\theta):=\ket{0}\bra{0} +e^{i\theta}\ket{1}\bra{1}. \label{Rtheta}\end{equation}
We furthermore define $T:=R\left(\pi/4\right)$ and the two-qubit SWAP gate
\begin{equation} \text{SWAP}:=\ket{00}\bra{00} + \ket{01}\bra{10} + \ket{10}\bra{01} +\ket{11}\bra{11}. \end{equation} 
Except for those gates defined above and the identity operator $\mathbb{I}$, where standard notation is used, all single-qubit gates will be denoted by lower case roman letters.
We define a general controlled gate, with a control qubit $j$ and a target qubit $k$, by
\begin{equation} C^j_{k}(u,v):=\ket{0}\bra{0}_j \otimes u_{k} + \ket{1}\bra{1}_j \otimes v_{k}, \end{equation} where $u,v \in U(2)$. The subscripts $j$ and $k$ will be dropped from the notation when no ambiguity will arise and we let $Cu:=C(\mathbb{I},u)$ and $SCu:=\text{SWAP} \cdot Cu$. Two operators $U,V \in U(4)$ are called \emph{locally equivalent} \cite{makhlin2002nonlocal} with respect to a decomposition into qubit subsystems $j$ and $k$ if
\begin{equation} U=u_j \otimes v_k \cdot V \cdot p_j \otimes q_k, \end{equation}
for some $u,v,p,q \in U(2)$.

\section{Minimal control ancilla-mediated quantum computation \label{min1}}
We now present two schemes of ancilla-mediated quantum computation that require only a single fixed-time ancilla-register interaction and ancilla preparation in the computational basis and hence are minimal control models.
\subsection{A first minimal control model}
We introduce our first model by giving a general form for an ancilla-register interaction which under certain conditions can implement a universal gate set on a register of qubits within the constraints of minimal control. We give an explicit construction for the application of a universal gate set on the register before comparing this model to the measurement-based scheme of Halil-Shah and Oi \cite{halil2014minimum} and giving a simple example of an interaction that obeys the required constraints. 
\subsubsection{A general interaction}
\begin{figure}[h!]
\center
\includegraphics[scale=1.2]{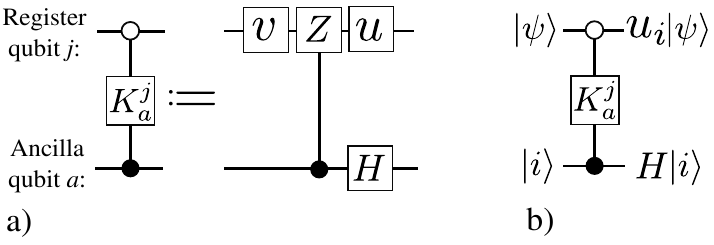}
\caption{\label{minint}a) The general $K^j_a$ ancilla-register interaction decomposed into local and non-local parts.  b) In this minimal control model the elements of a universal set for single-qubit unitaries, $\{u_0,u_1\}$, are applied to the register by the $K^j_{a}$ interaction and initialising the ancilla in the state $\ket{i}$ where $i=0,1$.}
\end{figure}

We consider a general fixed ancilla-register interaction of the form
\begin{equation} K^j_{a} :=  u_{j} \otimes H_a \cdot CZ \cdot  v_{j} \otimes \mathbb{I}_a,\end{equation}
where $u, v \in U(2)$. This interaction is shown in Fig.~\ref{minint}a and it is locally equivalent to $CZ$. We define $u_{0}:=uv$ and $u_1:=uZv$ and note that we may also write the interaction in the form $K^j_a=\mathbb{I} \otimes H \cdot C^a_j(u_0,u_1)$. We will show that for any $K$ such that $\{ u_0, u_1\}$ is a universal single-qubit gate set we may implement a minimal control model.
We do this by showing how we may implement a universal gate set on the register qubits. It follows directly from the definition of the ancilla-register interaction $K^j_a$ that
\begin{equation} K^j_{a} \ket{\psi}_j \ket{i}_a =  u_{i} \ket{\psi}_j \otimes H\ket{i}_a, \end{equation}
where $i=0,1$. Hence, we can deterministically apply the elements of $\{u_0,u_1\}$ on any register qubit, which we assume is a universal set for $SU(2)$, and so we may simulate any gate in $SU(2)$ up to arbitrary accuracy using only $K$ and the initialisation of ancilla in the computational basis. This gate method is depicted in the circuit diagram of Fig.~\ref{minint}b.
\newline
\indent
We now show how to implement a maximally entangling gate between two register qubits, $j$ and $k$, using only $K$ and ancillas prepared in the computational basis. A straightforward explicit calculation, utilising the identities $ vu^{\dagger}_0u= \mathbb{I}$,  $HZH=X$, $XX=ZZ=\mathbb{I}$ and $XZXZ=-\mathbb{I}$, shows that
\begin{equation} K^k_{a}  K^j_{a} \cdot u_{0_j}^{\dagger}    \otimes u_{0_k}^{\dagger} \otimes \mathbb{I} \cdot K^k_{a} K^j_{a}   = M^j_k \otimes \mathbb{I}_a, \label{minacqcseq} \end{equation}
where the induced entangling gate on the register qubits $j$ and $k$ is
\begin{equation} M^j_k= u_j \otimes u_k \cdot CZ \cdot v_j \otimes  v_k.\label{minM} \end{equation}
Although Eq.~(\ref{minacqcseq}) is an ancilla-mediated sequence which implements an entangling two-qubit gate on the register, as written it requires local unitaries on the register qubits and it is not decomposed into only $K$ gates. However, we may decompose the $u_0^{\dagger}$ gate on each register qubit into further $K$ gates. This is because $u_0$ and $u_1$ are a universal set for $SU(2)$ and hence there is a choice of $k_1,...,k_n=0,1$ such that 
\begin{equation}
\left(\prod_{i=1}^n K^{j}_{a_i}\right) \ket{\psi}_j\ket{k_{1}}_{a_1}...\ket{k_{n}}_{a_n}=\tilde{u}^{\dagger}_{0}\ket{\psi}_j H\ket{k_i}_{a_1}...H\ket{k_n}_{a_n},\end{equation}
with $\tilde{u}^{\dagger}_0=u_{k_n}...u_{k_1}$ approximating $u^{\dagger}_0$ up to arbitrary accuracy with finite $n$. In certain cases $u^{\dagger}_0$ may be implemented exactly. 
\begin{figure}[h!]
\center
\includegraphics[scale=1.2]{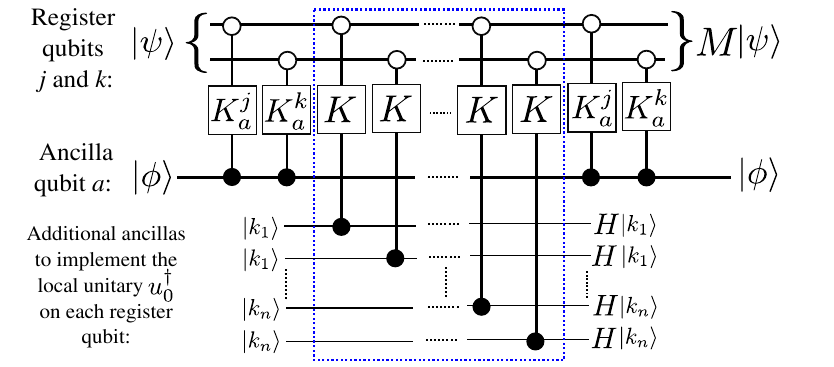}
\caption{ \label{mintwo} Two-qubit gates in the first minimal control model with interactions of the form $K^j_a$. This gate sequence has no overall effect on the `entangling' ancilla qubit but implements the entangling unitary $M$, given by Eq.~(\ref{minM}) and which is locally equivalent to $CZ$, on the register qubits. $2n$ additional ancillas with appropriate initial states in the computational basis are required to implement the unitary $u_0^{\dagger}$ (up to arbitrary accuracy) on each register qubit and these extra interactions are shown in the blue dashed box. The number of these additional ancillas depends on the form of $u_0$ and $u_1$. }
\end{figure}
Hence we may implement an entangling gate which is locally equivalent to $CZ$ between pairs of register qubits using only $K$ gates and ancillas initialised in the computational basis\textsuperscript{a}. The circuit diagram for this two-qubit gate on the register is given in Fig.~\ref{mintwo}. Therefore, under the assumption that $u_0$ and $u_1$ are universal for $SU(2)$, we have shown that $K$, along with ancillas prepared in the computational basis, can implement a minimal control model of ancilla-mediated quantum computation.
\subsubsection{Discussion and comparison with a measurement-based model}
We compare this minimal unitary model to the recently proposed measurement-based minimal scheme of Halil-Shah and Oi \cite{halil2014minimum}. In their model the computation is achieved using only a fixed interaction, ancilla preparation in a fixed state, and measurement in the computational basis. The fixed initial ancilla state is compensated for by the measurement which projects the ancilla onto states in the computational basis. Only one and two ancilla-register interactions are required to implement the single and two-qubit gates respectively. However this model results in a probabilistic repeat-until-success style gate scheme \cite{paetznick2013repeat}, whereby one has to wait until a random walk through the set of unitaries is within the required precision of the desired unitary. Although the model presented here is entirely deterministic, there is an overhead, that in general may be large, to implement the two-qubit gates. We note that this overhead does not appear if an ancilla measurement is performed after the first two interactions in Fig.~\ref{mintwo}, for an appropriately initialised ancilla and measurement basis, and in this case both models are of a similar form.
\newline
\indent
We note that if $vu$ is diagonal in the computational basis the additional ancillas are not required, however in this case $u_0$ and $u_1$ will commute and so cannot form a universal set for $SU(2)$. In general, the required additional ancillas for the two-qubit gates may create an impractical overhead. However, we now demonstrate that there exists an appropriate form for $K$ that is universal and has a low overhead for additional ancillas for each entangling gate.

\subsubsection{A specific interaction}
We can parametrise a general unitary operator $p \in U(2)$ by the matrix in the computational basis
\begin{equation}  p(\eta,\phi,\psi,\theta) = e^{i \eta}\begin{pmatrix} e^{i \phi}\cos \theta & e^{-i \psi} \sin \theta  \\  e^{i \psi}\sin \theta   & - e^{-i \phi} \cos \theta \end{pmatrix}. \end{equation} 
A specific suitable choice for the ancilla-register interaction $K$ is given by taking $u=p(\eta, \zeta,\zeta, \frac{\pi}{8})$ and $v=p(\frac{\pi}{8} - \eta, -\zeta-\frac{\pi}{8} ,\zeta - \frac{\pi}{8} , \frac{\pi}{8})$. It is straightforward to show that this gives $u_0 = T$ and $u_1=HT$. We have that $T^{7}=T^{\dagger}$ and so $u_1u_0^7=H$. It then follows that $u_0$ and $u_1$ form a universal set for $SU(2)$ as $H$ and $T$ are a universal set for single-qubit unitaries \cite{boykin2000new}. It is necessary to implement $u_0^{\dagger}$ on each register qubit to implement the sequence of Eq.~(\ref{minacqcseq}) and Fig.~\ref{mintwo}. We have that $u_0^{\dagger}=u_0^7$ and so the sequence of Eq.~(\ref{minacqcseq}) and Fig.~\ref{mintwo} can be implemented using 14 ancillas prepared in the state $\ket{0}$ and one `entangling' ancilla, that mediates the gate, prepared in any state.

%
\subsection{A second minimal control model \label{min2}}
We now present an alternative minimal control model which does not require additional ancillas. As before, we will introduce the model with a general interaction which with certain restrictions can be used to implement a minimal control model and then give a specific example of a simple suitable interaction.  
\subsubsection{A general interaction}
\begin{figure}[h!]
\center
\includegraphics[scale=1.2]{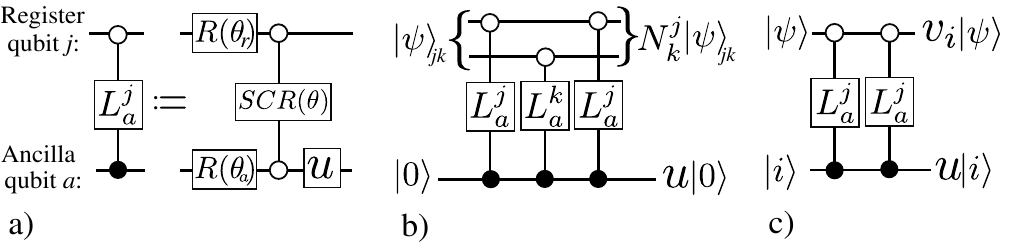}
\caption{ \label{minacqc} a) The decomposition of the fixed ancilla-register interaction $L^j_a$ into local and non-local parts. b) The two-qubit entangling gate $N^j_k$ is implemented on the register qubits $j$ and $k$ via the sequence of Eq.~(\ref{2qmin}) with an ancilla initialised in the state $\ket{0}$, where the form of $N^j_k$ is given in Eq.~(\ref{njk}). This two-qubit gate is not symmetric with respect to the exchange of $j$ and $k$. c) The single-qubit gate $v_i=R(\theta^i+\theta_a)uR(\theta^i+\theta_r)$ is applied to a register qubit $j$ by two applications of $L$ with an ancilla initialised in the state $\ket{i}$, $i=0,1$.}
\end{figure}

Take an ancilla-register interaction of the form
\begin{equation} L^j_{a} :=   \mathbb{I}_j \otimes  u_a \cdot SCR(\theta) \cdot R(\theta_r)_j \otimes R(\theta_a)_a. \end{equation}
This is decomposed into local and non-local parts in Fig.~\ref{minacqc}a. We note that this can also be expressed as $L^j_a=SC^a_j(uR(\theta_r),uR(\theta+\theta_r))\cdot\mathbb{I}_j \otimes R(\theta_a)_a$. We show that an interaction of this form, along with ancillas prepared in the computational basis, can implement universal quantum computation on the register if $\theta$ is such that $CR(\theta)$ is entangling (all non-trivial $\theta$) and $\{v_0,v_1\}$ is a universal set for $SU(2)$ where $v_i:=R(\theta^i+\theta_a)uR(\theta^i+\theta_r) $. As before, we do this by showing how we may implement a two-qubit entangling gate and a universal set for single-qubit unitaries on the register. We note that it is possible to set $\theta_r$ and $\theta_a$ to zero and obtain a universal interaction and these local rotations are included to increase the generality of the interaction.  
\newline
\indent
We may implement an entangling two-qubit gate between register qubits $j$ and $k$ using an ancilla initialised in the state $\ket{0}$ by interacting the ancilla sequentially with qubits $j$ and $k$ before completing the gate with a second interaction with the $j$ qubit. This is the interaction sequence
\begin{equation}L^j_{a}  L^k_{a}L^j_{a}  \ket{\psi}_{jk} \ket{0}_a =  N^j_k \ket{\psi}_{jk} \otimes u\ket{0}_a ,\label{2qmin}\end{equation}
where $N^j_k$ is entangling for non-trivial $\theta$ and is given by
\begin{equation} N^j_k = R(\theta_a)u_j \otimes \mathbb{I}_k \cdot SCR(\theta) \cdot R(\theta_a)uR(\theta_r)_j \otimes R(\theta_r)_k. \label{njk}\end{equation}
This is represented in the circuit diagram of Fig.~\ref{minacqc}b and can be shown with a simple explicit calculation.
We may decompose any single-qubit gate on a register qubit into only ancilla-register interactions $L$ and ancilla state-preparation in the computational basis. This is because
\begin{equation} L^j_aL^j_a  \ket{\psi}_j \ket{i}_a =  v_i\ket{\psi}_j \otimes u\ket{i}_a,\end{equation}
where $i=0,1$ and we assume that $\{v_0,v_1\}$ is a universal set for $SU(2)$. This is represented in the circuit diagram of Fig.~\ref{minacqc}c. Hence, as we have shown how to implement a two-qubit entangling gate and a universal set for $SU(2)$ on the register then this is a minimal control model of ancilla-mediated quantum computation.
\subsubsection{Discussion and comparison with other models}
This model requires three interactions for each entangling two-qubit gate on the register which, although greater than the two needed with the aid of ancilla measurement in ADQC \cite{anders2010ancilla} and the minimal extension of Halil-Shah and Oi \cite{halil2014minimum}, is the minimum possible in any measurement-free scheme \cite{lamata2008sequential}. Furthermore, in contrast to the first minimal control model, there is no requirement for multiple ancilla qubits to implement the entangling gates. Finally, we note that the two-qubit gates on the register are implemented in an identical fashion to those in the ACQC model \cite{proctor2013universal} and $L$ obeys the required conditions to be universal for that model. 
\subsubsection{A specific interaction}
A simple example of a specific form for the interaction $L^j_a$ such that $v_0$ and $v_1$ form a universal set for $SU(2)$ (and hence $L^j_a$ may implement this minimal control model) is given by taking $u=H$, $\theta=\frac{\pi}{4}$ and $\theta_r=\theta_a=0$. We then have that $L^j_{a} =   \mathbb{I}_j \otimes  H_a \cdot SCT$ and hence
  $v_0=H$ and $v_1=THT$. A proof of the universality of $\{H,THT\}$ for $SU(2)$ is given in Appendix~\ref{AppA}. The entangling gate induced on a pair of register qubits from the sequence of Eq.~(\ref{2qmin}) is then $N= H \otimes \mathbb{I} \cdot SCT \cdot H \otimes \mathbb{I}$ and this can easily simulate $\text{CNOT}:=CX$ as in this case $(N^j_k)^4=C^k_jX$.

\subsection{The physical implementation of minimal control models}
The models introduced herein are motivated by the physical challenges of implementing the multiple forms of control required in most universal models of quantum computation and hence it is interesting to briefly consider systems that may be appropriate for a physical realisation. We initially concentrate on a simple Hamiltonian for implementing the second minimal control model presented in Section~\ref{min2}. We set $\hbar=1$ and consider the two-qubit interaction Hamiltonian
\begin{equation} \boldsymbol{H}(\theta) = \pi (X \otimes X + Y \otimes Y) + (\pi - \theta) Z \otimes Z, \label{ham} \end{equation}
which, applied for a time $t=1/4$, implements (up to an irrelevant global phase) the unitary operator
\begin{equation} U(\theta)=e^{- i \boldsymbol{H}(\theta) /4} = SCR(\theta) \cdot R(-\theta/2) \otimes R(-\theta/2). \end{equation} 
If we consider the second minimal control model and take the fixed ancilla-register gate $L^j_a$ to be the unitary implemented by applying $\boldsymbol{H}(\theta)$ to the ancilla and register qubit for a time $t=1/4$, i.e. $U(\theta)$, followed by a fixed ancilla rotation of the form $R(\theta/2)HR(\theta/2)$, we have that $L^j_a = \mathbb{I}_j \otimes R(\theta/2)HR(\theta/2)_a \cdot U(\theta)$. Hence, this $L^j_a$ gives $v_0=H$ and $v_1=R(\theta)HR(\theta)$ which we have shown to be a universal set for $SU(2)$ when $\theta=\pi/4$ and so this form for $L^j_a$ is appropriate for implementing the second minimal control model. With this simple interaction Hamiltonian, $\boldsymbol{H}(\pi/4)$, local control of the ancilla is required. However, we see that this is a \emph{fixed} gate on the ancilla after every ancilla-register interaction via $\boldsymbol{H}(\pi/4)$ and hence this can be a fixed element in an experimental setup or incorporated into the natural evolution of the ancilla between interactions\textsuperscript{b}. For example, if the ancillary qubit is photonic the local operation can be performed by fixed linear optics \cite{kok2007linear} after each ancilla-register interaction. Indeed, the use of ancillary photons to mediate gates has been demonstrated in many experimental setups, for example with atomic \cite{reiserer2014quantum,tiecke2014nanophotonic} or spin \cite{carter2013quantum,luxmoore2013interfacing} qubits.
\newline
\indent
Interactions with the form $\boldsymbol{H}(\theta)$ arise naturally in spin systems, with one example of an implementation of $\boldsymbol{H}(\pi/4)$ given by the coupling between quantum dot resonant exchange qubits \cite{doherty2013two}. A particularly relevant physical system to ancilla-mediated models is the coupling of nuclear spins via ancillary electronic spins in nitrogen-vacancy (NV) defects in diamond \cite{taminiau2014universal,robledo2011high,waldherr2014quantum} and in such setups it may be possible to engineer the Hamiltonian $\boldsymbol{H}(\pi/4)$ \cite{borhani2005cluster}. Although in some physical realisations, such as the photonic case discussed above, the fixed local operation on the ancillary qubit after each interaction is convenient or natural, in others it may be problematic and negate the benefits of the models introduced herein. However, it is also possible to find Hamiltonians that directly implement suitable interactions for either of the models proposed in Sections~\ref{min1}. In certain systems the Hamiltonian is highly tuneable, with an example being those involving superconducting qubits \cite{niskanen2007quantum}, and due to the long coherence times of ensembles of electron spins in NV centers  \cite{stanwix2010coherence} a particularly promising physical system for ancilla-mediated quantum computing is arrays of spin ensembles in diamond (each spin ensemble is an effective qubit) coupled by ancillary flux qubits \cite{qiu2014coupling,lu2013quantum,marcos2010coupling}. Indeed, coherent coupling in such a system as been demonstrated \cite{zhu2011coherent}. It would be interesting to consider which physical systems have Hamiltonians that are naturally suited to generating appropriate interactions for the models introduced herein and we leave a more detailed study of this for future work.

\section{Conclusions \label{con}}
We have presented two unitary models of ancilla-mediated quantum computation that require only minimal control of both the ancillary and register systems. The only control necessary in these models to implement universal quantum computation on a register of qubits is a single fixed-time ancilla-register interaction between one ancilla qubit and one register qubit (at a time) and ancilla preparation in the computational basis. The first of these models is based on maximally entangling interactions that are locally equivalent to $CZ$ and requires multiple ancilla qubits to mediate two-qubit entangling gates on the register. This model is similar in many respects to the minimal measurement-based ancilla-mediated model of Halil-Shah and Oi \cite{halil2014minimum}, in which the requirement for ancilla preparation is replaced with the need for ancilla measurements in the computational basis, but is deterministic rather than stochastic. The second of these models removes the need for multiple ancillas to mediate each entangling gate by employing interactions that utilise the SWAP gate in a similar manner to the model known as ancilla-controlled quantum computation \cite{proctor2013universal}. As in the ancilla-controlled model, only three ancilla-register interactions are required to implement a two-qubit entangling gate on the register, which is the minimum possible in any scheme that does not include measurements \cite{lamata2008sequential}, and two for a single-qubit gate. Again, due to the global unitarity of the model the computation is deterministic and is based on a finite gate set composed of one two-qubit entangling gate and two single-qubit gates that form a universal set for single-qubit unitaries. We conjecture that these models require the minimal possible level of control for a unitary ancilla-mediated scheme.


\begin{backmatter}

\section*{Competing interests}
  The authors declare that they have no competing interests.
\section*{Acknowledgements}
TJP was supported by a university of Leeds Research Scholarship.
\section*{Endnotes}
\textsuperscript{a} The two-qubit gate in this first minimal control model can be seen to employ essentially the same geometric methods as the qudit ancilla model we introduced in Ref. \cite{proctor2014quantum} and hence also the qubus model. In the qudit ancilla model the ancilla interacts with the register via displacements of the ancilla (with the displacements defined in the discrete lattice phase space of a qudit) controlled by a register qubit. In this model with a qudit of dimension $2$, i.e. a qubit ancilla, controlled displacements are the $C^j_aX$ and $C^j_aZ$ gates. The two-qubit gates between register qubits $j$ and $k$ are then mediated by a sequence of the form $C^k_aXC^j_aZC^k_aXC^j_aZ=C^j_kZ$. This requires two different interactions between the ancilla and register. The model presented here essential uses this gate method but removes the need for two different interactions by including a Hadamard gate on the ancilla in the interaction definition as $HZH=X$. It is the additional local gates $u_l$ and $u_r$ (needed to make the gate universal without additional single-qubit gates) that then results in the need for additional ancillas to mediate the two-qubit gates in the first minimal control model. Finally, this gate method can be considered to be geometric as $XZXZ$ can be considered to create a closed loop in the discrete lattice phase space of a qubit - the details of this are in Ref. \cite{proctor2014quantum}.
\newline
\textsuperscript{b} Note that this is different to ADQC and ACQC in which the required rotations on the ancilla depend on the gate that is to be implemented (and previous measurement outcomes in the case of ADQC).

\bibliographystyle{bmc-mathphys} 
\bibliography{MyLibrary}      


\begin{thebibliography}{46}
\ifx \bisbn   \undefined \def \bisbn  #1{ISBN #1}\fi
\ifx \binits  \undefined \def \binits#1{#1}\fi
\ifx \bauthor  \undefined \def \bauthor#1{#1}\fi
\ifx \batitle  \undefined \def \batitle#1{#1}\fi
\ifx \bjtitle  \undefined \def \bjtitle#1{#1}\fi
\ifx \bvolume  \undefined \def \bvolume#1{\textbf{#1}}\fi
\ifx \byear  \undefined \def \byear#1{#1}\fi
\ifx \bissue  \undefined \def \bissue#1{#1}\fi
\ifx \bfpage  \undefined \def \bfpage#1{#1}\fi
\ifx \blpage  \undefined \def \blpage #1{#1}\fi
\ifx \burl  \undefined \def \burl#1{\textsf{#1}}\fi
\ifx \doiurl  \undefined \def \doiurl#1{\textsf{#1}}\fi
\ifx \betal  \undefined \def \betal{\textit{et al.}}\fi
\ifx \binstitute  \undefined \def \binstitute#1{#1}\fi
\ifx \binstitutionaled  \undefined \def \binstitutionaled#1{#1}\fi
\ifx \bctitle  \undefined \def \bctitle#1{#1}\fi
\ifx \beditor  \undefined \def \beditor#1{#1}\fi
\ifx \bpublisher  \undefined \def \bpublisher#1{#1}\fi
\ifx \bbtitle  \undefined \def \bbtitle#1{#1}\fi
\ifx \bedition  \undefined \def \bedition#1{#1}\fi
\ifx \bseriesno  \undefined \def \bseriesno#1{#1}\fi
\ifx \blocation  \undefined \def \blocation#1{#1}\fi
\ifx \bsertitle  \undefined \def \bsertitle#1{#1}\fi
\ifx \bsnm \undefined \def \bsnm#1{#1}\fi
\ifx \bsuffix \undefined \def \bsuffix#1{#1}\fi
\ifx \bparticle \undefined \def \bparticle#1{#1}\fi
\ifx \barticle \undefined \def \barticle#1{#1}\fi
\ifx \bconfdate \undefined \def \bconfdate #1{#1}\fi
\ifx \botherref \undefined \def \botherref #1{#1}\fi
\ifx \url \undefined \def \url#1{\textsf{#1}}\fi
\ifx \bchapter \undefined \def \bchapter#1{#1}\fi
\ifx \bbook \undefined \def \bbook#1{#1}\fi
\ifx \bcomment \undefined \def \bcomment#1{#1}\fi
\ifx \oauthor \undefined \def \oauthor#1{#1}\fi
\ifx \citeauthoryear \undefined \def \citeauthoryear#1{#1}\fi
\ifx \endbibitem  \undefined \def \endbibitem {}\fi
\ifx \bconflocation  \undefined \def \bconflocation#1{#1}\fi
\ifx \arxivurl  \undefined \def \arxivurl#1{\textsf{#1}}\fi
\csname PreBibitemsHook\endcsname

\bibitem{feynman1985quantum}
\begin{barticle}
\bauthor{\bsnm{Feynman}, \binits{R.}}:
\batitle{Quantum mechanical computers}.
\bjtitle{Optics news}
\bvolume{11}(\bissue{11-20}),
\bfpage{1}
(\byear{1985})
\end{barticle}
\endbibitem

\bibitem{brylinski2002universal}
\begin{bbook}
\bauthor{\bsnm{Brylinski}, \binits{J.-L.}},
\bauthor{\bsnm{Brylinski}, \binits{R.}}:
\bbtitle{Universal Quantum Gates},
pp. \bfpage{101}--\blpage{116}.
\bpublisher{Chapman \& Hall / CRC Press}
(\byear{2002})
\end{bbook}
\endbibitem

\bibitem{bremner2002practical}
\begin{barticle}
\bauthor{\bsnm{Bremner}, \binits{M.J.}},
\bauthor{\bsnm{Dawson}, \binits{C.M.}},
\bauthor{\bsnm{Dodd}, \binits{J.L.}},
\bauthor{\bsnm{Gilchrist}, \binits{A.}},
\bauthor{\bsnm{Harrow}, \binits{A.W.}},
\bauthor{\bsnm{Mortimer}, \binits{D.}},
\bauthor{\bsnm{Nielsen}, \binits{M.A.}},
\bauthor{\bsnm{Osborne}, \binits{T.J.}}:
\batitle{Practical scheme for quantum computation with any two-qubit entangling
  gate}.
\bjtitle{Phys. Rev. Lett.}
\bvolume{89}(\bissue{24}),
\bfpage{247902}
(\byear{2002})
\end{barticle}
\endbibitem

\bibitem{lloyd1995almost}
\begin{barticle}
\bauthor{\bsnm{Lloyd}, \binits{S.}}:
\batitle{Almost any quantum logic gate is universal}.
\bjtitle{Phys. Rev. Lett.}
\bvolume{75}(\bissue{2}),
\bfpage{346}
(\byear{1995})
\end{barticle}
\endbibitem

\bibitem{barenco1995universal}
\begin{barticle}
\bauthor{\bsnm{Barenco}, \binits{A.}}:
\batitle{A universal two-bit gate for quantum computation}.
\bjtitle{Proceedings of the Royal Society of London. Series A: Mathematical and
  Physical Sciences}
\bvolume{449}(\bissue{1937}),
\bfpage{679}--\blpage{683}
(\byear{1995})
\end{barticle}
\endbibitem

\bibitem{cirac1995quantum}
\begin{barticle}
\bauthor{\bsnm{Cirac}, \binits{J.I.}},
\bauthor{\bsnm{Zoller}, \binits{P.}}:
\batitle{Quantum computations with cold trapped ions}.
\bjtitle{Phys. Rev. Lett.}
\bvolume{74},
\bfpage{4091}
(\byear{1995})
\end{barticle}
\endbibitem

\bibitem{marcos2010coupling}
\begin{barticle}
\bauthor{\bsnm{Marcos}, \binits{D.}},
\bauthor{\bsnm{Wubs}, \binits{M.}},
\bauthor{\bsnm{Taylor}, \binits{J.M.}},
\bauthor{\bsnm{Aguado}, \binits{R.}},
\bauthor{\bsnm{Lukin}, \binits{M.D.}},
\bauthor{\bsnm{S{\o}rensen}, \binits{A.S.}}:
\batitle{Coupling nitrogen-vacancy centers in diamond to superconducting flux
  qubits}.
\bjtitle{Phys. Rev. Lett.}
\bvolume{105}(\bissue{21}),
\bfpage{210501}
(\byear{2010})
\end{barticle}
\endbibitem

\bibitem{lu2013quantum}
\begin{barticle}
\bauthor{\bsnm{L{\"u}}, \binits{X.-Y.}},
\bauthor{\bsnm{Xiang}, \binits{Z.-L.}},
\bauthor{\bsnm{Cui}, \binits{W.}},
\bauthor{\bsnm{You}, \binits{J.Q.}},
\bauthor{\bsnm{Nori}, \binits{F.}}:
\batitle{Quantum memory using a hybrid circuit with flux qubits and
  nitrogen-vacancy centers}.
\bjtitle{Phys. Rev. A}
\bvolume{88}(\bissue{1}),
\bfpage{012329}
(\byear{2013})
\end{barticle}
\endbibitem

\bibitem{qiu2014coupling}
\begin{barticle}
\bauthor{\bsnm{Qiu}, \binits{Y.}},
\bauthor{\bsnm{Xiong}, \binits{W.}},
\bauthor{\bsnm{Tian}, \binits{L.}},
\bauthor{\bsnm{You}, \binits{J.Q.}}:
\batitle{Coupling spin ensembles via superconducting flux qubits}.
\bjtitle{Phys. Rev. A}
\bvolume{89}(\bissue{4}),
\bfpage{042321}
(\byear{2014})
\end{barticle}
\endbibitem

\bibitem{zhu2011coherent}
\begin{barticle}
\bauthor{\bsnm{Zhu}, \binits{X.}},
\bauthor{\bsnm{Saito}, \binits{S.}},
\bauthor{\bsnm{Kemp}, \binits{A.}},
\bauthor{\bsnm{Kakuyanagi}, \binits{K.}},
\bauthor{\bsnm{Karimoto}, \binits{S.-i.}},
\bauthor{\bsnm{Nakano}, \binits{H.}},
\bauthor{\bsnm{Munro}, \binits{W.J.}},
\bauthor{\bsnm{Tokura}, \binits{Y.}},
\bauthor{\bsnm{Everitt}, \binits{M.S.}},
\bauthor{\bsnm{Nemoto}, \binits{K.}}, \betal:
\batitle{Coherent coupling of a superconducting flux qubit to an electron spin
  ensemble in diamond}.
\bjtitle{Nature}
\bvolume{478}(\bissue{7368}),
\bfpage{221}--\blpage{224}
(\byear{2011})
\end{barticle}
\endbibitem

\bibitem{wang2009coupling}
\begin{barticle}
\bauthor{\bsnm{Wang}, \binits{Y.-D.}},
\bauthor{\bsnm{Kemp}, \binits{A.}},
\bauthor{\bsnm{Semba}, \binits{K.}}:
\batitle{Coupling superconducting flux qubits at optimal point via dynamic
  decoupling with the quantum bus}.
\bjtitle{Phys. Rev. B}
\bvolume{79}(\bissue{2}),
\bfpage{024502}
(\byear{2009})
\end{barticle}
\endbibitem

\bibitem{wallraff2004strong}
\begin{barticle}
\bauthor{\bsnm{Wallraff}, \binits{A.}},
\bauthor{\bsnm{Schuster}, \binits{D.I.}},
\bauthor{\bsnm{Blais}, \binits{A.}},
\bauthor{\bsnm{Frunzio}, \binits{L.}},
\bauthor{\bsnm{Huang}, \binits{R.-S.}},
\bauthor{\bsnm{Majer}, \binits{J.}},
\bauthor{\bsnm{Kumar}, \binits{S.}},
\bauthor{\bsnm{Girvin}, \binits{S.M.}},
\bauthor{\bsnm{Schoelkopf}, \binits{R.J.}}:
\batitle{Strong coupling of a single photon to a superconducting qubit using
  circuit quantum electrodynamics}.
\bjtitle{Nature}
\bvolume{431}(\bissue{7005}),
\bfpage{162}--\blpage{167}
(\byear{2004})
\end{barticle}
\endbibitem

\bibitem{carter2013quantum}
\begin{barticle}
\bauthor{\bsnm{Carter}, \binits{S.G.}},
\bauthor{\bsnm{Sweeney}, \binits{T.M.}},
\bauthor{\bsnm{Kim}, \binits{M.}},
\bauthor{\bsnm{Kim}, \binits{C.S.}},
\bauthor{\bsnm{Solenov}, \binits{D.}},
\bauthor{\bsnm{Economou}, \binits{S.E.}},
\bauthor{\bsnm{Reinecke}, \binits{T.L.}},
\bauthor{\bsnm{Yang}, \binits{L.}},
\bauthor{\bsnm{Bracker}, \binits{A.S.}},
\bauthor{\bsnm{Gammon}, \binits{D.}}:
\batitle{Quantum control of a spin qubit coupled to a photonic crystal cavity}.
\bjtitle{Nature Photonics}
\bvolume{7}(\bissue{4}),
\bfpage{329}--\blpage{334}
(\byear{2013})
\end{barticle}
\endbibitem

\bibitem{luxmoore2013interfacing}
\begin{barticle}
\bauthor{\bsnm{Luxmoore}, \binits{I.}},
\bauthor{\bsnm{Wasley}, \binits{N.}},
\bauthor{\bsnm{Ramsay}, \binits{A.}},
\bauthor{\bsnm{Thijssen}, \binits{A.}},
\bauthor{\bsnm{Oulton}, \binits{R.}},
\bauthor{\bsnm{Hugues}, \binits{M.}},
\bauthor{\bsnm{Kasture}, \binits{S.}},
\bauthor{\bsnm{Achanta}, \binits{V.}},
\bauthor{\bsnm{Fox}, \binits{A.}},
\bauthor{\bsnm{Skolnick}, \binits{M.}}:
\batitle{Interfacing spins in an ingaas quantum dot to a semiconductor
  waveguide circuit using emitted photons}.
\bjtitle{Phys. Rev. Lett.}
\bvolume{110}(\bissue{3}),
\bfpage{037402}
(\byear{2013})
\end{barticle}
\endbibitem

\bibitem{armour2002entanglement}
\begin{barticle}
\bauthor{\bsnm{Armour}, \binits{A.D.}},
\bauthor{\bsnm{Blencowe}, \binits{M.P.}},
\bauthor{\bsnm{Schwab}, \binits{K.C.}}:
\batitle{Entanglement and decoherence of a micromechanical resonator via
  coupling to a cooper-pair box}.
\bjtitle{Phys. Rev. Lett.}
\bvolume{88}(\bissue{14}),
\bfpage{148301}--\blpage{148301}
(\byear{2002})
\end{barticle}
\endbibitem

\bibitem{milburn1999simulating}
\begin{botherref}
\oauthor{\bsnm{Milburn}, \binits{G.J.}}:
Simulating nonlinear spin models in an ion trap.
arXiv:quant-ph/9908037
(1999)
\end{botherref}
\endbibitem

\bibitem{wang2002simulation}
\begin{barticle}
\bauthor{\bsnm{Wang}, \binits{X.}},
\bauthor{\bsnm{Zanardi}, \binits{P.}}:
\batitle{Simulation of many-body interactions by conditional geometric phases}.
\bjtitle{Phys. Rev. A}
\bvolume{65}(\bissue{3}),
\bfpage{032327}
(\byear{2002})
\end{barticle}
\endbibitem

\bibitem{spiller2006quantum}
\begin{barticle}
\bauthor{\bsnm{Spiller}, \binits{T.P.}},
\bauthor{\bsnm{Nemoto}, \binits{K.}},
\bauthor{\bsnm{Braunstein}, \binits{S.L.}},
\bauthor{\bsnm{Munro}, \binits{W.J.}},
\bauthor{\bparticle{van} \bsnm{Loock}, \binits{P.}},
\bauthor{\bsnm{Milburn}, \binits{G.J.}}:
\batitle{Quantum computation by communication}.
\bjtitle{New J. Phys.}
\bvolume{8}(\bissue{2}),
\bfpage{30}
(\byear{2006})
\end{barticle}
\endbibitem

\bibitem{brown2011ancilla}
\begin{barticle}
\bauthor{\bsnm{Brown}, \binits{K.L.}},
\bauthor{\bsnm{De}, \binits{S.}},
\bauthor{\bsnm{Kendon}, \binits{V.M.}},
\bauthor{\bsnm{Munro}, \binits{W.J.}}:
\batitle{Ancilla-based quantum simulation}.
\bjtitle{New J. Phys.}
\bvolume{13}(\bissue{9}),
\bfpage{095007}
(\byear{2011})
\end{barticle}
\endbibitem

\bibitem{louis2007efficiencies}
\begin{barticle}
\bauthor{\bsnm{Louis}, \binits{S.G.R.}},
\bauthor{\bsnm{Nemoto}, \binits{K.}},
\bauthor{\bsnm{Munro}, \binits{W.J.}},
\bauthor{\bsnm{Spiller}, \binits{T.P.}}:
\batitle{The efficiencies of generating cluster states with weak
  nonlinearities}.
\bjtitle{New J. Phys.}
\bvolume{9}(\bissue{6}),
\bfpage{193}
(\byear{2007})
\end{barticle}
\endbibitem

\bibitem{munro2005efficient}
\begin{barticle}
\bauthor{\bsnm{Munro}, \binits{W.J.}},
\bauthor{\bsnm{Nemoto}, \binits{K.}},
\bauthor{\bsnm{Spiller}, \binits{T.P.}},
\bauthor{\bsnm{Barrett}, \binits{S.D.}},
\bauthor{\bsnm{Kok}, \binits{P.}},
\bauthor{\bsnm{Beausoleil}, \binits{R.G.}}:
\batitle{Efficient optical quantum information processing}.
\bjtitle{J. Opt. B: Quantum Semiclass. Opt.}
\bvolume{7}(\bissue{7}),
\bfpage{135}
(\byear{2005})
\end{barticle}
\endbibitem

\bibitem{proctor2014quantum}
\begin{botherref}
\oauthor{\bsnm{Proctor}, \binits{T.J.}},
\oauthor{\bsnm{Dooley}, \binits{S.}},
\oauthor{\bsnm{Kendon}, \binits{V.}}:
Quantum computation mediated by ancillary qudits and spin coherent states.
arXiv preprint arXiv:1402.6674v3
(2014)
\end{botherref}
\endbibitem

\bibitem{wootters1987wigner}
\begin{barticle}
\bauthor{\bsnm{Wootters}, \binits{W.K.}}:
\batitle{A wigner-function formulation of finite-state quantum mechanics}.
\bjtitle{Ann. Phys.}
\bvolume{176}(\bissue{1}),
\bfpage{1}--\blpage{21}
(\byear{1987})
\end{barticle}
\endbibitem

\bibitem{vourdas2004quantum}
\begin{barticle}
\bauthor{\bsnm{Vourdas}, \binits{A.}}:
\batitle{Quantum systems with finite hilbert space}.
\bjtitle{Rep. Prog. Phys.}
\bvolume{67}(\bissue{3}),
\bfpage{267}
(\byear{2004})
\end{barticle}
\endbibitem

\bibitem{noteE}
\begin{botherref}
Unpublished work in progress.
\end{botherref}
\endbibitem

\bibitem{brown2011using}
\begin{bbook}
\bauthor{\bsnm{Brown}, \binits{K.L.}}:
\bbtitle{Using the Qubus for Quantum Computing}.
\bpublisher{PhD Thesis, University of Leeds}
(\byear{2011})
\end{bbook}
\endbibitem

\bibitem{kashefi2009twisted}
\begin{barticle}
\bauthor{\bsnm{Kashefi}, \binits{E.}},
\bauthor{\bsnm{Oi}, \binits{D.K.L.}},
\bauthor{\bsnm{Browne}, \binits{D.}},
\bauthor{\bsnm{Anders}, \binits{J.}},
\bauthor{\bsnm{Andersson}, \binits{E.}}:
\batitle{Twisted graph states for ancilla-driven universal quantum
  computation}.
\bjtitle{Electronic Notes in Theoretical Computer Science}
\bvolume{249},
\bfpage{307}--\blpage{331}
(\byear{2009})
\end{barticle}
\endbibitem

\bibitem{anders2010ancilla}
\begin{barticle}
\bauthor{\bsnm{Anders}, \binits{J.}},
\bauthor{\bsnm{Oi}, \binits{D.K.L.}},
\bauthor{\bsnm{Kashefi}, \binits{E.}},
\bauthor{\bsnm{Browne}, \binits{D.E.}},
\bauthor{\bsnm{Andersson}, \binits{E.}}:
\batitle{Ancilla-driven universal quantum computation}.
\bjtitle{Phys. Rev. A}
\bvolume{82}(\bissue{2}),
\bfpage{020301}
(\byear{2010})
\end{barticle}
\endbibitem

\bibitem{shah2013ancilla}
\begin{bchapter}
\bauthor{\bsnm{Halil~Shah}, \binits{K.}},
\bauthor{\bsnm{Oi}, \binits{D.K.L.}}:
\bctitle{Ancilla Driven Quantum Computation with Arbitrary Entangling
  Strength}.
In: \bbtitle{Theory of Quantum Computation, Communication, and Cryptography,
  8th Conference, TQC 2013, LIPIcs-Leibniz International Proceedings in
  Informatics, Vol. 23.}
(\byear{2013})
\end{bchapter}
\endbibitem

\bibitem{proctor2013universal}
\begin{barticle}
\bauthor{\bsnm{Proctor}, \binits{T.J.}},
\bauthor{\bsnm{Andersson}, \binits{E.}},
\bauthor{\bsnm{Kendon}, \binits{V.}}:
\batitle{Universal quantum computation by the unitary control of ancilla qubits
  and using a fixed ancilla-register interaction}.
\bjtitle{Phys. Rev. A}
\bvolume{88}(\bissue{4}),
\bfpage{042330}
(\byear{2013})
\end{barticle}
\endbibitem

\bibitem{halil2014minimum}
\begin{botherref}
\oauthor{\bsnm{Halil-Shah}, \binits{K.}},
\oauthor{\bsnm{Oi}, \binits{D.K.L.}}:
A minimum control ancilla driven quantum computation scheme with
  repeat-until-success style gate generation.
arXiv preprint arXiv:1401.8004
(2014)
\end{botherref}
\endbibitem

\bibitem{paetznick2013repeat}
\begin{botherref}
\oauthor{\bsnm{Paetznick}, \binits{A.}},
\oauthor{\bsnm{Svore}, \binits{K.M.}}:
Repeat-until-success: Non-deterministic decomposition of single-qubit
  unitaries.
arXiv preprint arXiv:1311.1074
(2013)
\end{botherref}
\endbibitem

\bibitem{lamata2008sequential}
\begin{barticle}
\bauthor{\bsnm{Lamata}, \binits{L.}},
\bauthor{\bsnm{Le{\'o}n}, \binits{J.}},
\bauthor{\bsnm{P{\'e}rez-Garc{\'\i}a}, \binits{D.}},
\bauthor{\bsnm{Salgado}, \binits{D.}},
\bauthor{\bsnm{Solano}, \binits{E.}}:
\batitle{Sequential implementation of global quantum operations}.
\bjtitle{Phys. Rev. Lett.}
\bvolume{101}(\bissue{18}),
\bfpage{180506}
(\byear{2008})
\end{barticle}
\endbibitem

\bibitem{makhlin2002nonlocal}
\begin{barticle}
\bauthor{\bsnm{Makhlin}, \binits{Y.}}:
\batitle{Nonlocal properties of two-qubit gates and mixed states, and the
  optimization of quantum computations}.
\bjtitle{Quantum Inf. Process.}
\bvolume{1}(\bissue{4}),
\bfpage{243}--\blpage{252}
(\byear{2002})
\end{barticle}
\endbibitem

\bibitem{boykin2000new}
\begin{barticle}
\bauthor{\bsnm{Boykin}, \binits{P.O.}},
\bauthor{\bsnm{Mor}, \binits{T.}},
\bauthor{\bsnm{Pulver}, \binits{M.}},
\bauthor{\bsnm{Roychowdhury}, \binits{V.}},
\bauthor{\bsnm{Vatan}, \binits{F.}}:
\batitle{A new universal and fault-tolerant quantum basis}.
\bjtitle{Inform. Process. Lett.}
\bvolume{75}(\bissue{3}),
\bfpage{101}--\blpage{107}
(\byear{2000})
\end{barticle}
\endbibitem

\bibitem{kok2007linear}
\begin{barticle}
\bauthor{\bsnm{Kok}, \binits{P.}},
\bauthor{\bsnm{Munro}, \binits{W.J.}},
\bauthor{\bsnm{Nemoto}, \binits{K.}},
\bauthor{\bsnm{Ralph}, \binits{T.C.}},
\bauthor{\bsnm{Dowling}, \binits{J.P.}},
\bauthor{\bsnm{Milburn}, \binits{G.J.}}:
\batitle{Linear optical quantum computing with photonic qubits}.
\bjtitle{Rev. Mod. Phys.}
\bvolume{79}(\bissue{1}),
\bfpage{135}
(\byear{2007})
\end{barticle}
\endbibitem

\bibitem{reiserer2014quantum}
\begin{barticle}
\bauthor{\bsnm{Reiserer}, \binits{A.}},
\bauthor{\bsnm{Kalb}, \binits{N.}},
\bauthor{\bsnm{Rempe}, \binits{G.}},
\bauthor{\bsnm{Ritter}, \binits{S.}}:
\batitle{A quantum gate between a flying optical photon and a single trapped
  atom}.
\bjtitle{Nature}
\bvolume{508}(\bissue{7495}),
\bfpage{237}--\blpage{240}
(\byear{2014})
\end{barticle}
\endbibitem

\bibitem{tiecke2014nanophotonic}
\begin{barticle}
\bauthor{\bsnm{Tiecke}, \binits{T.}},
\bauthor{\bsnm{Thompson}, \binits{J.}},
\bauthor{\bparticle{de} \bsnm{Leon}, \binits{N.}},
\bauthor{\bsnm{Liu}, \binits{L.}},
\bauthor{\bsnm{Vuleti{\'c}}, \binits{V.}},
\bauthor{\bsnm{Lukin}, \binits{M.}}:
\batitle{Nanophotonic quantum phase switch with a single atom}.
\bjtitle{Nature}
\bvolume{508}(\bissue{7495}),
\bfpage{241}--\blpage{244}
(\byear{2014})
\end{barticle}
\endbibitem

\bibitem{doherty2013two}
\begin{barticle}
\bauthor{\bsnm{Doherty}, \binits{A.C.}},
\bauthor{\bsnm{Wardrop}, \binits{M.P.}}:
\batitle{Two-qubit gates for resonant exchange qubits}.
\bjtitle{Phys. Rev. Lett.}
\bvolume{111}(\bissue{5}),
\bfpage{050503}
(\byear{2013})
\end{barticle}
\endbibitem

\bibitem{taminiau2014universal}
\begin{botherref}
\oauthor{\bsnm{Taminiau}, \binits{T.H.}},
\oauthor{\bsnm{Cramer}, \binits{J.}},
\oauthor{\bparticle{van~der} \bsnm{Sar}, \binits{T.}},
\oauthor{\bsnm{Dobrovitski}, \binits{V.V.}},
\oauthor{\bsnm{Hanson}, \binits{R.}}:
Universal control and error correction in multi-qubit spin registers in
  diamond.
Nat. nanotechnol.
(2014)
\end{botherref}
\endbibitem

\bibitem{robledo2011high}
\begin{barticle}
\bauthor{\bsnm{Robledo}, \binits{L.}},
\bauthor{\bsnm{Childress}, \binits{L.}},
\bauthor{\bsnm{Bernien}, \binits{H.}},
\bauthor{\bsnm{Hensen}, \binits{B.}},
\bauthor{\bsnm{Alkemade}, \binits{P.F.A.}},
\bauthor{\bsnm{Hanson}, \binits{R.}}:
\batitle{High-fidelity projective read-out of a solid-state spin quantum
  register}.
\bjtitle{Nature}
\bvolume{477}(\bissue{7366}),
\bfpage{574}--\blpage{578}
(\byear{2011})
\end{barticle}
\endbibitem

\bibitem{waldherr2014quantum}
\begin{barticle}
\bauthor{\bsnm{Waldherr}, \binits{G.}},
\bauthor{\bsnm{Wang}, \binits{Y.}},
\bauthor{\bsnm{Zaiser}, \binits{S.}},
\bauthor{\bsnm{Jamali}, \binits{M.}},
\bauthor{\bsnm{Schulte-Herbr{\"u}ggen}, \binits{T.}},
\bauthor{\bsnm{Abe}, \binits{H.}},
\bauthor{\bsnm{Ohshima}, \binits{T.}},
\bauthor{\bsnm{Isoya}, \binits{J.}},
\bauthor{\bsnm{Du}, \binits{J.F.}},
\bauthor{\bsnm{Neumann}, \binits{P.}},
\bauthor{\bsnm{Wrachtrup}, \binits{J.}}:
\batitle{Quantum error correction in a solid-state hybrid spin register}.
\bjtitle{Nature}
\bvolume{506}(\bissue{7487}),
\bfpage{204}--\blpage{207}
(\byear{2014})
\end{barticle}
\endbibitem

\bibitem{borhani2005cluster}
\begin{barticle}
\bauthor{\bsnm{Borhani}, \binits{M.}},
\bauthor{\bsnm{Loss}, \binits{D.}}:
\batitle{Cluster states from heisenberg interactions}.
\bjtitle{Phys. Rev. A}
\bvolume{71}(\bissue{3}),
\bfpage{034308}
(\byear{2005})
\end{barticle}
\endbibitem

\bibitem{niskanen2007quantum}
\begin{barticle}
\bauthor{\bsnm{Niskanen}, \binits{A.O.}},
\bauthor{\bsnm{Harrabi}, \binits{K.}},
\bauthor{\bsnm{Yoshihara}, \binits{F.}},
\bauthor{\bsnm{Nakamura}, \binits{Y.}},
\bauthor{\bsnm{Lloyd}, \binits{S.}},
\bauthor{\bsnm{Tsai}, \binits{J.S.}}:
\batitle{Quantum coherent tunable coupling of superconducting qubits}.
\bjtitle{Science}
\bvolume{316}(\bissue{5825}),
\bfpage{723}--\blpage{726}
(\byear{2007})
\end{barticle}
\endbibitem

\bibitem{stanwix2010coherence}
\begin{barticle}
\bauthor{\bsnm{Stanwix}, \binits{P.L.}},
\bauthor{\bsnm{Pham}, \binits{L.M.}},
\bauthor{\bsnm{Maze}, \binits{J.R.}},
\bauthor{\bsnm{Le~Sage}, \binits{D.}},
\bauthor{\bsnm{Yeung}, \binits{T.K.}},
\bauthor{\bsnm{Cappellaro}, \binits{P.}},
\bauthor{\bsnm{Hemmer}, \binits{P.R.}},
\bauthor{\bsnm{Yacoby}, \binits{A.}},
\bauthor{\bsnm{Lukin}, \binits{M.D.}},
\bauthor{\bsnm{Walsworth}, \binits{R.L.}}:
\batitle{Coherence of nitrogen-vacancy electronic spin ensembles in diamond}.
\bjtitle{Phys. Rev. B}
\bvolume{82}(\bissue{20}),
\bfpage{201201}
(\byear{2010})
\end{barticle}
\endbibitem

\bibitem{mladenova2011vector}
\begin{barticle}
\bauthor{\bsnm{Mladenova}, \binits{C.D.}},
\bauthor{\bsnm{Mladenov}, \binits{I.M.}}:
\batitle{Vector decomposition of finite rotations}.
\bjtitle{Rep. Math. Phys.}
\bvolume{68}(\bissue{1}),
\bfpage{107}--\blpage{117}
(\byear{2011})
\end{barticle}
\endbibitem

\end{thebibliography}

\newcommand{\BMCxmlcomment}[1]{}

\BMCxmlcomment{

<refgrp>

<bibl id="B1">
  <title><p>Quantum mechanical computers</p></title>
  <aug>
    <au><snm>Feynman</snm><fnm>R.</fnm></au>
  </aug>
  <source>Optics news</source>
  <pubdate>1985</pubdate>
  <volume>11</volume>
  <issue>11-20</issue>
  <fpage>1</fpage>
</bibl>

<bibl id="B2">
  <title><p>Universal quantum gates</p></title>
  <aug>
    <au><snm>Brylinski</snm><fnm>J. L.</fnm></au>
    <au><snm>Brylinski</snm><fnm>R.</fnm></au>
  </aug>
  <source>Mathematics of Quantum Computation</source>
  <publisher>Chapman \& Hall / CRC Press</publisher>
  <pubdate>2002</pubdate>
  <fpage>101</fpage>
  <lpage>116</lpage>
</bibl>

<bibl id="B3">
  <title><p>Practical scheme for quantum computation with any two-qubit
  entangling gate</p></title>
  <aug>
    <au><snm>Bremner</snm><fnm>M. J.</fnm></au>
    <au><snm>Dawson</snm><fnm>C. M.</fnm></au>
    <au><snm>Dodd</snm><fnm>J. L.</fnm></au>
    <au><snm>Gilchrist</snm><fnm>A.</fnm></au>
    <au><snm>Harrow</snm><fnm>A. W.</fnm></au>
    <au><snm>Mortimer</snm><fnm>D.</fnm></au>
    <au><snm>Nielsen</snm><fnm>M. A.</fnm></au>
    <au><snm>Osborne</snm><fnm>T. J.</fnm></au>
  </aug>
  <source>Phys. Rev. Lett.</source>
  <publisher>APS</publisher>
  <pubdate>2002</pubdate>
  <volume>89</volume>
  <issue>24</issue>
  <fpage>247902</fpage>
</bibl>

<bibl id="B4">
  <title><p>Almost any quantum logic gate is universal</p></title>
  <aug>
    <au><snm>Lloyd</snm><fnm>S.</fnm></au>
  </aug>
  <source>Phys. Rev. Lett.</source>
  <publisher>APS</publisher>
  <pubdate>1995</pubdate>
  <volume>75</volume>
  <issue>2</issue>
  <fpage>346</fpage>
</bibl>

<bibl id="B5">
  <title><p>A universal two-bit gate for quantum computation</p></title>
  <aug>
    <au><snm>Barenco</snm><fnm>A.</fnm></au>
  </aug>
  <source>Proceedings of the Royal Society of London. Series A: Mathematical
  and Physical Sciences</source>
  <publisher>The Royal Society</publisher>
  <pubdate>1995</pubdate>
  <volume>449</volume>
  <issue>1937</issue>
  <fpage>679</fpage>
  <lpage>-683</lpage>
</bibl>

<bibl id="B6">
  <title><p>Quantum computations with cold trapped ions</p></title>
  <aug>
    <au><snm>Cirac</snm><fnm>J. I.</fnm></au>
    <au><snm>Zoller</snm><fnm>P.</fnm></au>
  </aug>
  <source>Phys. Rev. Lett.</source>
  <pubdate>1995</pubdate>
  <volume>74</volume>
  <fpage>4091</fpage>
</bibl>

<bibl id="B7">
  <title><p>Coupling nitrogen-vacancy centers in diamond to superconducting
  flux qubits</p></title>
  <aug>
    <au><snm>Marcos</snm><fnm>D.</fnm></au>
    <au><snm>Wubs</snm><fnm>M.</fnm></au>
    <au><snm>Taylor</snm><fnm>J. M.</fnm></au>
    <au><snm>Aguado</snm><fnm>R.</fnm></au>
    <au><snm>Lukin</snm><fnm>M. D.</fnm></au>
    <au><snm>S{\o}rensen</snm><fnm>A. S.</fnm></au>
  </aug>
  <source>Phys. Rev. Lett.</source>
  <publisher>APS</publisher>
  <pubdate>2010</pubdate>
  <volume>105</volume>
  <issue>21</issue>
  <fpage>210501</fpage>
</bibl>

<bibl id="B8">
  <title><p>Quantum memory using a hybrid circuit with flux qubits and
  nitrogen-vacancy centers</p></title>
  <aug>
    <au><snm>L{\"u}</snm><fnm>X. Y.</fnm></au>
    <au><snm>Xiang</snm><fnm>Z. L.</fnm></au>
    <au><snm>Cui</snm><fnm>W.</fnm></au>
    <au><snm>You</snm><fnm>J. Q.</fnm></au>
    <au><snm>Nori</snm><fnm>F.</fnm></au>
  </aug>
  <source>Phys. Rev. A</source>
  <publisher>APS</publisher>
  <pubdate>2013</pubdate>
  <volume>88</volume>
  <issue>1</issue>
  <fpage>012329</fpage>
</bibl>

<bibl id="B9">
  <title><p>Coupling spin ensembles via superconducting flux qubits</p></title>
  <aug>
    <au><snm>Qiu</snm><fnm>Y.</fnm></au>
    <au><snm>Xiong</snm><fnm>W.</fnm></au>
    <au><snm>Tian</snm><fnm>L.</fnm></au>
    <au><snm>You</snm><fnm>J. Q.</fnm></au>
  </aug>
  <source>Phys. Rev. A</source>
  <publisher>APS</publisher>
  <pubdate>2014</pubdate>
  <volume>89</volume>
  <issue>4</issue>
  <fpage>042321</fpage>
</bibl>

<bibl id="B10">
  <title><p>Coherent coupling of a superconducting flux qubit to an electron
  spin ensemble in diamond</p></title>
  <aug>
    <au><snm>Zhu</snm><fnm>X.</fnm></au>
    <au><snm>Saito</snm><fnm>S.</fnm></au>
    <au><snm>Kemp</snm><fnm>A.</fnm></au>
    <au><snm>Kakuyanagi</snm><fnm>K.</fnm></au>
    <au><snm>Karimoto</snm><fnm>Si</fnm></au>
    <au><snm>Nakano</snm><fnm>H.</fnm></au>
    <au><snm>Munro</snm><fnm>W. J.</fnm></au>
    <au><snm>Tokura</snm><fnm>Y.</fnm></au>
    <au><snm>Everitt</snm><fnm>M. S.</fnm></au>
    <au><snm>Nemoto</snm><fnm>K.</fnm></au>
    <au><cnm>others</cnm></au>
  </aug>
  <source>Nature</source>
  <publisher>Nature Publishing Group</publisher>
  <pubdate>2011</pubdate>
  <volume>478</volume>
  <issue>7368</issue>
  <fpage>221</fpage>
  <lpage>-224</lpage>
</bibl>

<bibl id="B11">
  <title><p>Coupling superconducting flux qubits at optimal point via dynamic
  decoupling with the quantum bus</p></title>
  <aug>
    <au><snm>Wang</snm><fnm>Y. D.</fnm></au>
    <au><snm>Kemp</snm><fnm>A.</fnm></au>
    <au><snm>Semba</snm><fnm>K.</fnm></au>
  </aug>
  <source>Phys. Rev. B</source>
  <publisher>APS</publisher>
  <pubdate>2009</pubdate>
  <volume>79</volume>
  <issue>2</issue>
  <fpage>024502</fpage>
</bibl>

<bibl id="B12">
  <title><p>Strong coupling of a single photon to a superconducting qubit using
  circuit quantum electrodynamics</p></title>
  <aug>
    <au><snm>Wallraff</snm><fnm>A.</fnm></au>
    <au><snm>Schuster</snm><fnm>D. I.</fnm></au>
    <au><snm>Blais</snm><fnm>A.</fnm></au>
    <au><snm>Frunzio</snm><fnm>L.</fnm></au>
    <au><snm>Huang</snm><fnm>R. S.</fnm></au>
    <au><snm>Majer</snm><fnm>J.</fnm></au>
    <au><snm>Kumar</snm><fnm>S.</fnm></au>
    <au><snm>Girvin</snm><fnm>S. M.</fnm></au>
    <au><snm>Schoelkopf</snm><fnm>R. J.</fnm></au>
  </aug>
  <source>Nature</source>
  <publisher>Nature Publishing Group</publisher>
  <pubdate>2004</pubdate>
  <volume>431</volume>
  <issue>7005</issue>
  <fpage>162</fpage>
  <lpage>-167</lpage>
</bibl>

<bibl id="B13">
  <title><p>Quantum control of a spin qubit coupled to a photonic crystal
  cavity</p></title>
  <aug>
    <au><snm>Carter</snm><fnm>S. G.</fnm></au>
    <au><snm>Sweeney</snm><fnm>T. M.</fnm></au>
    <au><snm>Kim</snm><fnm>M.</fnm></au>
    <au><snm>Kim</snm><fnm>C. S.</fnm></au>
    <au><snm>Solenov</snm><fnm>D.</fnm></au>
    <au><snm>Economou</snm><fnm>S. E</fnm></au>
    <au><snm>Reinecke</snm><fnm>T. L.</fnm></au>
    <au><snm>Yang</snm><fnm>L.</fnm></au>
    <au><snm>Bracker</snm><fnm>A. S.</fnm></au>
    <au><snm>Gammon</snm><fnm>D.</fnm></au>
  </aug>
  <source>Nature Photonics</source>
  <publisher>Nature Publishing Group</publisher>
  <pubdate>2013</pubdate>
  <volume>7</volume>
  <issue>4</issue>
  <fpage>329</fpage>
  <lpage>-334</lpage>
</bibl>

<bibl id="B14">
  <title><p>Interfacing spins in an InGaAs quantum dot to a semiconductor
  waveguide circuit using emitted photons</p></title>
  <aug>
    <au><snm>Luxmoore</snm><fnm>IJ</fnm></au>
    <au><snm>Wasley</snm><fnm>NA</fnm></au>
    <au><snm>Ramsay</snm><fnm>AJ</fnm></au>
    <au><snm>Thijssen</snm><fnm>ACT</fnm></au>
    <au><snm>Oulton</snm><fnm>R</fnm></au>
    <au><snm>Hugues</snm><fnm>M</fnm></au>
    <au><snm>Kasture</snm><fnm>S</fnm></au>
    <au><snm>Achanta</snm><fnm>VG</fnm></au>
    <au><snm>Fox</snm><fnm>AM</fnm></au>
    <au><snm>Skolnick</snm><fnm>MS</fnm></au>
  </aug>
  <source>Phys. Rev. Lett.</source>
  <publisher>APS</publisher>
  <pubdate>2013</pubdate>
  <volume>110</volume>
  <issue>3</issue>
  <fpage>037402</fpage>
</bibl>

<bibl id="B15">
  <title><p>Entanglement and decoherence of a micromechanical resonator via
  coupling to a Cooper-pair box</p></title>
  <aug>
    <au><snm>Armour</snm><fnm>A. D.</fnm></au>
    <au><snm>Blencowe</snm><fnm>M. P.</fnm></au>
    <au><snm>Schwab</snm><fnm>K. C.</fnm></au>
  </aug>
  <source>Phys. Rev. Lett.</source>
  <publisher>American Physical Society</publisher>
  <pubdate>2002</pubdate>
  <volume>88</volume>
  <issue>14</issue>
  <fpage>148301</fpage>
  <lpage>-148301</lpage>
</bibl>

<bibl id="B16">
  <title><p>Simulating nonlinear spin models in an ion trap</p></title>
  <aug>
    <au><snm>Milburn</snm><fnm>G. J.</fnm></au>
  </aug>
  <source>arXiv:quant-ph/9908037</source>
  <pubdate>1999</pubdate>
</bibl>

<bibl id="B17">
  <title><p>Simulation of many-body interactions by conditional geometric
  phases</p></title>
  <aug>
    <au><snm>Wang</snm><fnm>X.</fnm></au>
    <au><snm>Zanardi</snm><fnm>P.</fnm></au>
  </aug>
  <source>Phys. Rev. A</source>
  <publisher>APS</publisher>
  <pubdate>2002</pubdate>
  <volume>65</volume>
  <issue>3</issue>
  <fpage>032327</fpage>
</bibl>

<bibl id="B18">
  <title><p>Quantum computation by communication</p></title>
  <aug>
    <au><snm>Spiller</snm><fnm>T. P.</fnm></au>
    <au><snm>Nemoto</snm><fnm>K.</fnm></au>
    <au><snm>Braunstein</snm><fnm>S. L.</fnm></au>
    <au><snm>Munro</snm><fnm>W. J.</fnm></au>
    <au><snm>Loock</snm><fnm>P.</fnm></au>
    <au><snm>Milburn</snm><fnm>G. J.</fnm></au>
  </aug>
  <source>New J. Phys.</source>
  <publisher>IOP Publishing</publisher>
  <pubdate>2006</pubdate>
  <volume>8</volume>
  <issue>2</issue>
  <fpage>30</fpage>
</bibl>

<bibl id="B19">
  <title><p>Ancilla-based quantum simulation</p></title>
  <aug>
    <au><snm>Brown</snm><fnm>K. L.</fnm></au>
    <au><snm>De</snm><fnm>S.</fnm></au>
    <au><snm>Kendon</snm><fnm>V. M.</fnm></au>
    <au><snm>Munro</snm><fnm>W. J.</fnm></au>
  </aug>
  <source>New J. Phys.</source>
  <publisher>IOP Publishing</publisher>
  <pubdate>2011</pubdate>
  <volume>13</volume>
  <issue>9</issue>
  <fpage>095007</fpage>
</bibl>

<bibl id="B20">
  <title><p>The efficiencies of generating cluster states with weak
  nonlinearities</p></title>
  <aug>
    <au><snm>Louis</snm><fnm>S. G. R.</fnm></au>
    <au><snm>Nemoto</snm><fnm>K.</fnm></au>
    <au><snm>Munro</snm><fnm>W. J.</fnm></au>
    <au><snm>Spiller</snm><fnm>T. P.</fnm></au>
  </aug>
  <source>New J. Phys.</source>
  <publisher>IOP Publishing</publisher>
  <pubdate>2007</pubdate>
  <volume>9</volume>
  <issue>6</issue>
  <fpage>193</fpage>
</bibl>

<bibl id="B21">
  <title><p>Efficient optical quantum information processing</p></title>
  <aug>
    <au><snm>Munro</snm><fnm>W. J.</fnm></au>
    <au><snm>Nemoto</snm><fnm>K.</fnm></au>
    <au><snm>Spiller</snm><fnm>T. P.</fnm></au>
    <au><snm>Barrett</snm><fnm>S. D.</fnm></au>
    <au><snm>Kok</snm><fnm>P.</fnm></au>
    <au><snm>Beausoleil</snm><fnm>R. G.</fnm></au>
  </aug>
  <source>J. Opt. B: Quantum Semiclass. Opt.</source>
  <publisher>IOP Publishing</publisher>
  <pubdate>2005</pubdate>
  <volume>7</volume>
  <issue>7</issue>
  <fpage>S135</fpage>
</bibl>

<bibl id="B22">
  <title><p>Quantum computation mediated by ancillary qudits and spin coherent
  states</p></title>
  <aug>
    <au><snm>Proctor</snm><fnm>T. J.</fnm></au>
    <au><snm>Dooley</snm><fnm>S.</fnm></au>
    <au><snm>Kendon</snm><fnm>V.</fnm></au>
  </aug>
  <source>arXiv preprint arXiv:1402.6674v3</source>
  <pubdate>2014</pubdate>
</bibl>

<bibl id="B23">
  <title><p>A Wigner-function formulation of finite-state quantum
  mechanics</p></title>
  <aug>
    <au><snm>Wootters</snm><fnm>W. K.</fnm></au>
  </aug>
  <source>Ann. Phys.</source>
  <publisher>Elsevier</publisher>
  <pubdate>1987</pubdate>
  <volume>176</volume>
  <issue>1</issue>
  <fpage>1</fpage>
  <lpage>-21</lpage>
</bibl>

<bibl id="B24">
  <title><p>Quantum systems with finite Hilbert space</p></title>
  <aug>
    <au><snm>Vourdas</snm><fnm>A.</fnm></au>
  </aug>
  <source>Rep. Prog. Phys.</source>
  <publisher>IOP Publishing</publisher>
  <pubdate>2004</pubdate>
  <volume>67</volume>
  <issue>3</issue>
  <fpage>267</fpage>
</bibl>

<bibl id="B25">
  <note>Unpublished work in progress.</note>
</bibl>

<bibl id="B26">
  <title><p>Using the qubus for quantum computing</p></title>
  <aug>
    <au><snm>Brown</snm><fnm>K. L.</fnm></au>
  </aug>
  <publisher>PhD Thesis, University of Leeds</publisher>
  <pubdate>2011</pubdate>
</bibl>

<bibl id="B27">
  <title><p>Twisted graph states for ancilla-driven universal quantum
  computation</p></title>
  <aug>
    <au><snm>Kashefi</snm><fnm>E.</fnm></au>
    <au><snm>Oi</snm><fnm>D. K. L.</fnm></au>
    <au><snm>Browne</snm><fnm>D.</fnm></au>
    <au><snm>Anders</snm><fnm>J.</fnm></au>
    <au><snm>Andersson</snm><fnm>E.</fnm></au>
  </aug>
  <source>Electronic Notes in Theoretical Computer Science</source>
  <publisher>Elsevier</publisher>
  <pubdate>2009</pubdate>
  <volume>249</volume>
  <fpage>307</fpage>
  <lpage>-331</lpage>
</bibl>

<bibl id="B28">
  <title><p>Ancilla-driven universal quantum computation</p></title>
  <aug>
    <au><snm>Anders</snm><fnm>J.</fnm></au>
    <au><snm>Oi</snm><fnm>D. K. L.</fnm></au>
    <au><snm>Kashefi</snm><fnm>E.</fnm></au>
    <au><snm>Browne</snm><fnm>D. E.</fnm></au>
    <au><snm>Andersson</snm><fnm>E.</fnm></au>
  </aug>
  <source>Phys. Rev. A</source>
  <publisher>APS</publisher>
  <pubdate>2010</pubdate>
  <volume>82</volume>
  <issue>2</issue>
  <fpage>020301</fpage>
</bibl>

<bibl id="B29">
  <title><p>Ancilla Driven Quantum Computation with arbitrary entangling
  strength</p></title>
  <aug>
    <au><snm>Halil Shah</snm><fnm>K.</fnm></au>
    <au><snm>Oi</snm><fnm>D. K. L.</fnm></au>
  </aug>
  <source>Theory of Quantum Computation, Communication, and Cryptography, 8th
  Conference, TQC 2013, LIPIcs-Leibniz International Proceedings in
  Informatics, Vol. 23.</source>
  <pubdate>2013</pubdate>
</bibl>

<bibl id="B30">
  <title><p>Universal quantum computation by the unitary control of ancilla
  qubits and using a fixed ancilla-register interaction</p></title>
  <aug>
    <au><snm>Proctor</snm><fnm>T. J.</fnm></au>
    <au><snm>Andersson</snm><fnm>E.</fnm></au>
    <au><snm>Kendon</snm><fnm>V.</fnm></au>
  </aug>
  <source>Phys. Rev. A</source>
  <publisher>APS</publisher>
  <pubdate>2013</pubdate>
  <volume>88</volume>
  <issue>4</issue>
  <fpage>042330</fpage>
</bibl>

<bibl id="B31">
  <title><p>A minimum control ancilla driven quantum computation scheme with
  repeat-until-success style gate generation</p></title>
  <aug>
    <au><snm>Halil Shah</snm><fnm>K.</fnm></au>
    <au><snm>Oi</snm><fnm>D. K. L.</fnm></au>
  </aug>
  <source>arXiv preprint arXiv:1401.8004</source>
  <pubdate>2014</pubdate>
</bibl>

<bibl id="B32">
  <title><p>Repeat-Until-Success: Non-deterministic decomposition of
  single-qubit unitaries</p></title>
  <aug>
    <au><snm>Paetznick</snm><fnm>A.</fnm></au>
    <au><snm>Svore</snm><fnm>K. M.</fnm></au>
  </aug>
  <source>arXiv preprint arXiv:1311.1074</source>
  <pubdate>2013</pubdate>
</bibl>

<bibl id="B33">
  <title><p>Sequential implementation of global quantum operations</p></title>
  <aug>
    <au><snm>Lamata</snm><fnm>L.</fnm></au>
    <au><snm>Le{\'o}n</snm><fnm>J.</fnm></au>
    <au><snm>P{\'e}rez Garc{\'\i}a</snm><fnm>D.</fnm></au>
    <au><snm>Salgado</snm><fnm>D.</fnm></au>
    <au><snm>Solano</snm><fnm>E.</fnm></au>
  </aug>
  <source>Phys. Rev. Lett.</source>
  <publisher>APS</publisher>
  <pubdate>2008</pubdate>
  <volume>101</volume>
  <issue>18</issue>
  <fpage>180506</fpage>
</bibl>

<bibl id="B34">
  <title><p>Nonlocal properties of two-qubit gates and mixed states, and the
  optimization of quantum computations</p></title>
  <aug>
    <au><snm>Makhlin</snm><fnm>Y.</fnm></au>
  </aug>
  <source>Quantum Inf. Process.</source>
  <publisher>Springer</publisher>
  <pubdate>2002</pubdate>
  <volume>1</volume>
  <issue>4</issue>
  <fpage>243</fpage>
  <lpage>-252</lpage>
</bibl>

<bibl id="B35">
  <title><p>A new universal and fault-tolerant quantum basis</p></title>
  <aug>
    <au><snm>Boykin</snm><fnm>P. O.</fnm></au>
    <au><snm>Mor</snm><fnm>T.</fnm></au>
    <au><snm>Pulver</snm><fnm>M.</fnm></au>
    <au><snm>Roychowdhury</snm><fnm>V.</fnm></au>
    <au><snm>Vatan</snm><fnm>F.</fnm></au>
  </aug>
  <source>Inform. Process. Lett.</source>
  <publisher>Elsevier</publisher>
  <pubdate>2000</pubdate>
  <volume>75</volume>
  <issue>3</issue>
  <fpage>101</fpage>
  <lpage>-107</lpage>
</bibl>

<bibl id="B36">
  <title><p>Linear optical quantum computing with photonic qubits</p></title>
  <aug>
    <au><snm>Kok</snm><fnm>P.</fnm></au>
    <au><snm>Munro</snm><fnm>W. J.</fnm></au>
    <au><snm>Nemoto</snm><fnm>K.</fnm></au>
    <au><snm>Ralph</snm><fnm>T. C.</fnm></au>
    <au><snm>Dowling</snm><fnm>J. P.</fnm></au>
    <au><snm>Milburn</snm><fnm>G. J.</fnm></au>
  </aug>
  <source>Rev. Mod. Phys.</source>
  <publisher>APS</publisher>
  <pubdate>2007</pubdate>
  <volume>79</volume>
  <issue>1</issue>
  <fpage>135</fpage>
</bibl>

<bibl id="B37">
  <title><p>A quantum gate between a flying optical photon and a single trapped
  atom</p></title>
  <aug>
    <au><snm>Reiserer</snm><fnm>A.</fnm></au>
    <au><snm>Kalb</snm><fnm>N.</fnm></au>
    <au><snm>Rempe</snm><fnm>G.</fnm></au>
    <au><snm>Ritter</snm><fnm>S.</fnm></au>
  </aug>
  <source>Nature</source>
  <publisher>Nature Publishing Group</publisher>
  <pubdate>2014</pubdate>
  <volume>508</volume>
  <issue>7495</issue>
  <fpage>237</fpage>
  <lpage>-240</lpage>
</bibl>

<bibl id="B38">
  <title><p>Nanophotonic quantum phase switch with a single atom</p></title>
  <aug>
    <au><snm>Tiecke</snm><fnm>TG</fnm></au>
    <au><snm>Thompson</snm><fnm>JD</fnm></au>
    <au><snm>Leon</snm><fnm>NP</fnm></au>
    <au><snm>Liu</snm><fnm>LR</fnm></au>
    <au><snm>Vuleti{\'c}</snm><fnm>V</fnm></au>
    <au><snm>Lukin</snm><fnm>MD</fnm></au>
  </aug>
  <source>Nature</source>
  <publisher>Nature Publishing Group</publisher>
  <pubdate>2014</pubdate>
  <volume>508</volume>
  <issue>7495</issue>
  <fpage>241</fpage>
  <lpage>-244</lpage>
</bibl>

<bibl id="B39">
  <title><p>Two-Qubit Gates for Resonant Exchange Qubits</p></title>
  <aug>
    <au><snm>Doherty</snm><fnm>A. C.</fnm></au>
    <au><snm>Wardrop</snm><fnm>M. P.</fnm></au>
  </aug>
  <source>Phys. Rev. Lett.</source>
  <publisher>APS</publisher>
  <pubdate>2013</pubdate>
  <volume>111</volume>
  <issue>5</issue>
  <fpage>050503</fpage>
</bibl>

<bibl id="B40">
  <title><p>Universal control and error correction in multi-qubit spin
  registers in diamond</p></title>
  <aug>
    <au><snm>Taminiau</snm><fnm>T. H.</fnm></au>
    <au><snm>Cramer</snm><fnm>J.</fnm></au>
    <au><snm>Sar</snm><fnm>T.</fnm></au>
    <au><snm>Dobrovitski</snm><fnm>V. V.</fnm></au>
    <au><snm>Hanson</snm><fnm>R.</fnm></au>
  </aug>
  <source>Nat. nanotechnol.</source>
  <publisher>Nature Publishing Group</publisher>
  <pubdate>2014</pubdate>
</bibl>

<bibl id="B41">
  <title><p>High-fidelity projective read-out of a solid-state spin quantum
  register</p></title>
  <aug>
    <au><snm>Robledo</snm><fnm>L.</fnm></au>
    <au><snm>Childress</snm><fnm>L.</fnm></au>
    <au><snm>Bernien</snm><fnm>H.</fnm></au>
    <au><snm>Hensen</snm><fnm>B.</fnm></au>
    <au><snm>Alkemade</snm><fnm>P. F. A.</fnm></au>
    <au><snm>Hanson</snm><fnm>R.</fnm></au>
  </aug>
  <source>Nature</source>
  <publisher>Nature Publishing Group</publisher>
  <pubdate>2011</pubdate>
  <volume>477</volume>
  <issue>7366</issue>
  <fpage>574</fpage>
  <lpage>-578</lpage>
</bibl>

<bibl id="B42">
  <title><p>Quantum error correction in a solid-state hybrid spin
  register</p></title>
  <aug>
    <au><snm>Waldherr</snm><fnm>G.</fnm></au>
    <au><snm>Wang</snm><fnm>Y.</fnm></au>
    <au><snm>Zaiser</snm><fnm>S.</fnm></au>
    <au><snm>Jamali</snm><fnm>M.</fnm></au>
    <au><snm>Schulte Herbr{\"u}ggen</snm><fnm>T.</fnm></au>
    <au><snm>Abe</snm><fnm>H.</fnm></au>
    <au><snm>Ohshima</snm><fnm>T.</fnm></au>
    <au><snm>Isoya</snm><fnm>J.</fnm></au>
    <au><snm>Du</snm><fnm>J. F.</fnm></au>
    <au><snm>Neumann</snm><fnm>P.</fnm></au>
    <au><snm>Wrachtrup</snm><fnm>J.</fnm></au>
  </aug>
  <source>Nature</source>
  <publisher>Nature Publishing Group</publisher>
  <pubdate>2014</pubdate>
  <volume>506</volume>
  <issue>7487</issue>
  <fpage>204</fpage>
  <lpage>-207</lpage>
</bibl>

<bibl id="B43">
  <title><p>Cluster states from Heisenberg interactions</p></title>
  <aug>
    <au><snm>Borhani</snm><fnm>M.</fnm></au>
    <au><snm>Loss</snm><fnm>D.</fnm></au>
  </aug>
  <source>Phys. Rev. A</source>
  <publisher>APS</publisher>
  <pubdate>2005</pubdate>
  <volume>71</volume>
  <issue>3</issue>
  <fpage>034308</fpage>
</bibl>

<bibl id="B44">
  <title><p>Quantum coherent tunable coupling of superconducting
  qubits</p></title>
  <aug>
    <au><snm>Niskanen</snm><fnm>A. O.</fnm></au>
    <au><snm>Harrabi</snm><fnm>K.</fnm></au>
    <au><snm>Yoshihara</snm><fnm>F.</fnm></au>
    <au><snm>Nakamura</snm><fnm>Y.</fnm></au>
    <au><snm>Lloyd</snm><fnm>S.</fnm></au>
    <au><snm>Tsai</snm><fnm>J. S.</fnm></au>
  </aug>
  <source>Science</source>
  <publisher>American Association for the Advancement of Science</publisher>
  <pubdate>2007</pubdate>
  <volume>316</volume>
  <issue>5825</issue>
  <fpage>723</fpage>
  <lpage>-726</lpage>
</bibl>

<bibl id="B45">
  <title><p>Coherence of nitrogen-vacancy electronic spin ensembles in
  diamond</p></title>
  <aug>
    <au><snm>Stanwix</snm><fnm>PL</fnm></au>
    <au><snm>Pham</snm><fnm>LM</fnm></au>
    <au><snm>Maze</snm><fnm>JR</fnm></au>
    <au><snm>Le Sage</snm><fnm>D</fnm></au>
    <au><snm>Yeung</snm><fnm>TK</fnm></au>
    <au><snm>Cappellaro</snm><fnm>P</fnm></au>
    <au><snm>Hemmer</snm><fnm>PR</fnm></au>
    <au><snm>Yacoby</snm><fnm>A</fnm></au>
    <au><snm>Lukin</snm><fnm>MD</fnm></au>
    <au><snm>Walsworth</snm><fnm>RL</fnm></au>
  </aug>
  <source>Phys. Rev. B</source>
  <publisher>APS</publisher>
  <pubdate>2010</pubdate>
  <volume>82</volume>
  <issue>20</issue>
  <fpage>201201</fpage>
</bibl>

<bibl id="B46">
  <title><p>Vector decomposition of finite rotations</p></title>
  <aug>
    <au><snm>Mladenova</snm><fnm>CD</fnm></au>
    <au><snm>Mladenov</snm><fnm>IM</fnm></au>
  </aug>
  <source>Rep. Math. Phys.</source>
  <publisher>Elsevier</publisher>
  <pubdate>2011</pubdate>
  <volume>68</volume>
  <issue>1</issue>
  <fpage>107</fpage>
  <lpage>-117</lpage>
</bibl>

</refgrp>
} 


\end{backmatter}

\appendix
\section{\label{AppA}}
Here we prove that $v_0=H$ and $v_1=THT$ are a universal set for $SU(2)$. Using similar notation to Boykin \emph{et al.} \cite{boykin2000new}, we denote the $n^{th}$ roots of the $X$ and $Z$ operators by $X^{\frac{1}{n}}$ and $Z^{\frac{1}{n}}$. Any $u \in SU(2)$ can be written as
\begin{equation} u=\exp \left( i \varphi \hat{n} \cdot \vec{\sigma} \right), \end{equation}
where $\vec{\sigma}=(X,Y,Z)$ is the vector of Pauli operators, $\hat{n}=(n_x,n_y,n_z)$ is some unit vector in $\mathbb{R}^3$, $\hat{n} \cdot \vec{\sigma} = n_x X + n_y Y +n_z Z$ and $\varphi \in \mathbb{R}$ is some rotation angle. We have that
\begin{equation} \exp \left( i \varphi \hat{n} \cdot \vec{\sigma} \right) = \cos \varphi \mathbb{I} + i \sin \varphi ( \hat{n} \cdot \vec{\sigma}  ).\end{equation}
Up to irrelevant global phases, which we ignore from now on, $Z=\exp \left( i \frac{\pi}{2} Z \right)$ and $X=\exp \left( i \frac{\pi}{2} X \right)$ and hence $Z^{\frac{1}{n}}=\exp \left( i \frac{\pi}{2n} Z \right)$ and $X^{\frac{1}{n}}=\exp \left( i \frac{\pi}{2n} X \right)$. Using these, and the identity $HZH=X$, we have that $X^{\frac{1}{n}}=HZ^{\frac{1}{n}}H$. It is straightforward to confirm that $T=Z^{-\frac{1}{4}}$ and so $v_+:=v_0v_1=X^{-\frac{1}{4}}Z^{-\frac{1}{4}}$ and $v_-:=v_1v_0=Z^{-\frac{1}{4}}X^{-\frac{1}{4}}$. From a simple explicit calculation, we have that
\begin{equation} v_{\pm} = \cos^2 \frac{\pi}{8} - i \sin^2 \frac{\pi}{8} \left(\cot \frac{\pi}{8}(Z+X) \mp Y\right). \end{equation}
Therefore, for both $v_+$ and $v_-$ we have that $\cos \varphi = \cos^2 \frac{\pi}{8}$ and hence $\varphi$ is an irrational multiple of $\pi$ \cite{boykin2000new} and $\hat{n}_{\pm}=n_{\pm}/\|n_{\pm}\|$ where $n_{\pm} = -(\cot\frac{\pi}{8},\mp1,\cot\frac{\pi}{8})$. As $\varphi$ is an irrational multiple of $\pi$ we can approximate to arbitrary accuracy any rotation around the $n_{\pm}$ axis by $m$ applications of $v_{\pm}$, with $m$ a finite integer. As these axes of rotation are not parallel then any arbitrary rotation can be decomposed into rotations around these axes \cite{mladenova2011vector}. This then proves that $v_+$ and $v_-$ and hence $v_0$ and $v_1$ are a universal set for $SU(2)$. 
\end{document}